\documentstyle[11pt,epsfig]{article}
\title{\bf Quantum noncommutative multidimensional cosmology}
\author{N. Khosravi\thanks{email:
n-khosravi@sbu.ac.ir}, S. Jalalzadeh\thanks{email:
s-jalalzadeh@sbu.ac.ir} and H. R. Sepangi\thanks{email:
hr-sepangi@sbu.ac.ir}
\\ {\small Department of Physics, Shahid Beheshti University, Evin,
Tehran 19839, Iran}}
\begin{document}
\maketitle

\begin{abstract}
We present exact quantum solutions for a noncommutative,
multidimensional cosmological model and show that stabilization of
extra dimensions sets in with the introduction of noncommutativity
between the scale factors. An interpretation is offered to
accommodate the notion of time, rendering  comparison with the
classical solutions possible.
\vspace{5mm}\noindent
\end{abstract}

\section{Introduction}
Noncommutativity between space-time coordinates, first introduced
in \cite{1}, has been a source of inspiration in recent years
\cite{2,3,4}. This interest has its roots in the development of
string and M-theories, \cite{5,6}.  In addition, interesting
predictions regarding, for example, the IR/UV mixing and
non-locality \cite{7}, Lorentz violation \cite{8} and new physics
at very short distance scales \cite{9,10,11} makes the study of
this subject worthwhile. The impact of noncommutativity in
cosmology has been considerable and addressed in different works
\cite{12}. Hopefully, noncommutative cosmology would lead us to
the formulation of semiclassical approximations of quantum gravity
and tackles the cosmological constant problem \cite{13}. Of
particular interest would be the application of noncommutativity
to multidimensional cosmology.

The oldest notion of multidimensional gravity dates back to the
now famous works of Kaluza and Klein \cite{kk}, where an extra
compact dimension was used to unify electromagnetism and gravity.
Since then, a large body of research work has been devoted to the
study of multidimensional models inspired by Kaluza and Klein. The
idea of extra dimensions have also been employed to suggest
possible resolutions to some of the important problems that
cosmology has been facing \cite{14}, namely, the mass hierarchy
and cosmological constant problem. As was mentioned above, over
the recent past, the notion of noncommutativity has been playing
an increasingly important role in cosmology and, naturally, the
Kaluza-Klein theory would have a prominent part to play in this
regard. An example of the works in which noncommutativity in the
context of Kaluza-Klein theory is used can be found in
\cite{general}.

As soon as extra dimensions are introduced, the question of their
compactification and stabilization becomes important and should be
addressed \cite{stab}. It would therefore be of interest to study
model theories where such issues are addressed and possible
solutions presented. In a previous work \cite{nima}, a
noncommutative multidimensional cosmological model was introduced
and used to address the issues of compactification and
stabilization of extra dimensions and the cosmological constant
problem at the classical regime.  In this letter, we study the
noncommutative quantum cosmology of this model by taking the scale
factors of our ordinary space and extra dimensions as
noncommutative variables and obtain exact solutions to the
Wheeler-DeWitt (WD) equation. The question of time, an important
issue in quantum cosmology, is addressed by interpreting the
wavefunctions representing the universe in a manner consistent
with the probabilistic interpretation of quantum mechanics.
\section{The Model}
We consider a cosmological model in which the space-time is
assumed to be of FRW type with a $d$-dimensional internal space.
The corresponding metric can be written as
\begin{equation}\label{1}
ds^2=-dt^2+\frac{R^2(t)}{\left(1+\frac{k}{4}r^2\right)^2}
(dr^2+r^2 d\Omega^2)+a^2(t)g^{(d)}_{ij}dx^idx^j
    \hspace{.15cm},
\end{equation}
where the total number of dimensions is $D=4+d$, $k=1,0,-1$
represents the usual spatial curvature, $R(t)$ and $a(t)$ are the
scale factors of the universe and the radius of the
$d$-dimensional internal space respectively and $g^{(d)}_{ij}$ is
the metric associated with the internal space, assumed to be
Ricci-flat. The Ricci scalar corresponding to metric (\ref{1}) is
\begin{equation}\label{2}
{\cal
R}=6\left[\frac{\ddot{R}}{R}+\frac{k+\dot{R}^{2}}{R^2}\right]+
2\,d\frac{\ddot{a}}{a}+d(d-1)\left(\frac{\dot{a}}{a}\right)^2\
+6\,d\frac{\dot{a}\dot{R}}{aR} ,
\end{equation}
where a dot represents differentiation with respect to $t$. Let
$a_0$ be the compactification scale of the internal space at the
present time and
\begin{equation}\label{3}
v_d\equiv v_0\times v_i\equiv
a_0^d\times\int_{M_d}d^dx\sqrt{-g^{(d)}},
\end{equation}
the corresponding total volume of the internal space. Substitution
of equation (\ref{2}) and use of definition (\ref{3}) in the
Einstein-Hilbert action functional with a $D$-dimensional
cosmological constant $\Lambda$
\begin{equation}\label{4}
{\cal{S}}=\frac{1}{2k_D^2}\int_M d^Dx \sqrt{-g}({\cal R}
-2{\Lambda})+{\cal{S}}_{YGH},
\end{equation}
where $k_D$ is the $D$-dimensional gravitational constant and
${\cal{S}}_{YGH}$ is the York-Gibbons-Hawking boundary term, leads
to, after dimensional reduction
\begin{equation}\label{5}
{\cal S}=-v_{D-1}\int dt \left\{6\dot{R}^2\Phi
R+6\dot{R}\dot{\Phi}R^2+\frac{d-1}{d}\frac{\dot{\Phi}^2}{\Phi}R^3-6k\Phi
R-2\Phi R^3\Lambda\right\},
\end{equation}
where
\begin{equation}\label{6}
\Phi=\left(\frac{a}{a_0}\right)^d,
\end{equation}
and we have set $v_{D-1}=1$. To make the Lagrangian manageable,
consider the following change of variables
\begin{equation}\label{8}
\Phi R^3=\Upsilon^2(x_1^2-x_2^2),
\end{equation}
where $R=R(x_1,x_2)$ and $\Phi=\Phi(x_1,x_2)$ are functions of new
variables $x_1$, $x_2$. Let
\begin{eqnarray}\label{9}
\left\{%
\begin{array}{lll}
\Phi^{\rho_+}R^{\sigma_-}=\Upsilon(x_1+x_2),\\
\\
\Phi^{\rho_-}R^{\sigma_+}=\Upsilon(x_1-x_2), \\
\end{array}%
\right.
\end{eqnarray}
such that for $d\neq3$, $\Upsilon=1$ and we have
\begin{eqnarray}\label{10}
\left\{%
\begin{array}{ll}
\rho_\pm=\frac{1}{2}\pm \frac{3}{4}\sqrt{\frac{d+2}{3d}}\mp
\frac{1}{4\sqrt{\frac{d+2}{3d}}},\\
\sigma_{\pm}=\frac{1}{2}\left(3\mp \frac{1}{\sqrt{\frac{d+2}{3d}}}\right),\\
\end{array}
\right.
\end{eqnarray}
while for $d=3$, $\Upsilon=\frac{3}{\sqrt{5}}$ and
\begin{eqnarray}\label{101}
\left\{%
\begin{array}{ll}
\rho_\pm=\frac{1}{2}\pm \frac{\sqrt{5}}{10},\\
\sigma_{\pm}=3\left(\frac{1}{2}\pm \frac{\sqrt{5}}{10}\right).\\
\end{array}
\right.
\end{eqnarray}
Using the above transformations introduced in \cite{shahram} and
concentrating on $k=0$, the Lagrangian becomes
\begin{equation}\label{11}
{\cal
L}=-4\left(\frac{d+2}{d+3}\right)\left\{\dot{x_1}^2-\dot{x_2}^2-
\frac{\Lambda}{2}\left(\frac{d+3}{d+2}\right)\left(x_1^2-x_2^2\right)\right\}.
\end{equation}
Up to an overall  constant coefficient, we can write the effective
Hamiltonian as
\begin{equation}\label{14}
{\cal
H}=\frac{p_1^2}{4}-\frac{p_2^2}{4}+\omega^2\left(x_1^2-x_2^2\right),
\end{equation}
where $\omega^2$ is
\begin{equation}\label{1a}
\omega^2=\frac{1}{2}\left(\frac{d+3}{d+2}\right)\Lambda.
\end{equation}
This Hamiltonian is known as that of an
oscillator-ghost-oscillators because of the wrong sign of $x_2$
\cite{17}. As it was shown in \cite{nima}, the solutions of the
equations of motion in this case result in the compactification of
the extra dimensions and suggests a possible resolution for the
cosmological constant problem. In the next section we study and
compare the quantum behavior of the commutative and noncommutative
solutions.
\section{Quantum solutions}
\subsection{Commutative case}
The transition from the classical setup to the quantum case starts
with replacing the phase space variables with quantum operators
such that
\begin{equation}\label{24}
\widehat{\cal
H}=\frac{1}{4}\eta^{\mu\nu}\hat{p}_\mu\hat{p}_\nu+\omega^2
\eta^{\mu\nu}\hat{x}_\mu\hat{x}_\nu,
\end{equation}
where
\begin{equation}\label{21}
[\hat{x}_\mu,\hat{p}_\nu]=i\delta_{{\mu}{\nu}},
\end{equation}
with $\hbar=1$ and we have noted that this is the only non-zero
commutator. The annihilation and creation operators are defined in
the usual way
\begin{equation}\label{22}
\begin{array}{lll}
a_\mu=\sqrt{\omega}\hat{x}_\mu+\frac{i}{2\sqrt{\omega}}\hat{p}_\mu,\\
\\
a_\mu^\dag=\sqrt{\omega}\hat{x}_\mu-\frac{i}{2\sqrt{\omega}}\hat{p}_\mu.
\end{array}
\end{equation}
These operators satisfy the following commutation relations
\begin{equation}\label{23}
\left[a_\mu,{a_{\nu}}^{\dag}\right]=\delta_{\mu\nu}
\hspace{.15cm}.
\end{equation}
The Hamiltonian operator  can readily be written in terms of
$a_\mu$ and $a_{\mu}^\dag$
\begin{equation}\label{25}
\begin{array}{lll}
{\widehat{\cal H}}&=&\frac{\omega}{2}\left(a_\mu^\dag a^\mu+a^\mu
a_\mu^\dag\right),\\
\\
&=&\omega\left({a_1}^\dag a_1-{a_2}^\dag a_2\right),\\
 \end{array}
\end{equation}
where in the second line, relation (\ref{23}) is used. We thus
define the vacuum states as
\begin{equation}\label{261}
\begin{array}{lll}
|0,0\rangle&=&|0\rangle_1\otimes|0\rangle_2,\\
a_1|0\rangle_1&=&0,\\
{a_2}|0\rangle_2&=&0,\\
\end{array}
\end{equation}
with the excited states given by
\begin{equation}\label{263}
\begin{array}{lll}
|n,m\rangle&=&|n\rangle_1\otimes|m\rangle_2,\\
|n\rangle_1&=&\frac{1}{\sqrt{n!}}({a_1}^\dag)^n|0\rangle_1,\\
|m\rangle_2&=&\frac{1}{\sqrt{m!}}({a_2}^\dag)^m|0\rangle_2,\\
\end{array}
\end{equation}
where $n$ and $m$ are integers. These states satisfy the WD
equation when $m=n$, that is
\begin{equation}\label{26}
\widehat{\cal H}|2n\rangle=0,
\end{equation}
where $|2n\rangle=|n,n\rangle$. The ground state is $|0,0\rangle$
or, in the $x$-representation
\begin{equation}\label{2611}
\Psi_{0}(x_1,x_2)={\cal N}_1
H_0(x_1)H_0(x_2)e^{-\omega(x_1^2+x_2^2) },
\end{equation}
where ${\cal N}_1$ is a normalization factor and $H_n(x)$ is the
appropriate Hermite polynomial. The other solutions of the WD
equation can be built from this solution by the action of the
creation operators such that
\begin{equation}\label{2621}
\Psi_{2n}(x_1,x_2)={\cal N}_2
H_n(x_1)H_{n}(x_2)e^{-\omega(x_1^2+x_2^2)},
\end{equation}
where ${\cal N}_2$ is a normalization factor. One can construct
the general solution, $\Psi(x_1,x_2)$, as the linear combinations
of the above solutions as
\begin{equation}\label{2631}
\Psi(x_1,x_2)={\cal N}\sum_{n=0}^{\infty}c_{n} \Psi_{2n}(x_1,x_2),
\end{equation}
where again ${\cal N}$ is a normalization constant. Coefficients
$c_n$ are chosen in such a way as to make the states coherent
\cite{18}, that is
\begin{equation}\label{2641}
c_n=e^{-\frac{1}{4}|\chi_0|^2}\frac{\chi_0^n}{\sqrt{2^n n!}},
\end{equation}
with $\chi_0$ being an arbitrary complex number. Figure 1 shows
the square of the wave function and the corresponding contour
plots.

\subsection{Noncommutative case}
In this section we endeavor to find the solutions for the
noncommutative oscillator-ghost-oscillator studied above. To
proceed, one has to define creation and annihilation-like
operators which commute with the Hamiltonian and therefore
generate the solutions. In what follows, we construct these
operators which are built out of non-linear combinations of the
usual creation and annihilation operators.

To study noncommutativity we start with equation (\ref{24}) and
write it with minimal variation as
\begin{equation}\label{50}
\widehat{\cal
H'}=\frac{1}{4}\eta^{\mu\nu}\hat{p'}_\mu\hat{p'}_\nu+
\omega^2\eta^{\mu\nu}\hat{x'}_\mu\hat{x'}_\nu,
\end{equation}
so that
\begin{eqnarray}\label{51}
\left[\hat{x'}_\mu,\hat{x'}_\nu\right]=\theta{\epsilon_{\mu\nu}},\hspace{4mm}
\left[\hat{x'}_\mu,\hat{p'}_\nu\right]=(1+\sigma)\delta_{\mu\nu},\hspace{4mm}
\left[\hat{p'}_\mu,\hat{p'}_\nu\right]=\beta {\epsilon_{\mu\nu}},
\end{eqnarray}
where $\sigma=\frac{\theta \beta}{4}$. Using the standard trick
\cite{sheikh}, we have
\begin{eqnarray}\label{511}
\hat{x'}_{\mu} = \hat{x}_{\mu}-\frac{1}{2}\theta
\epsilon_{\mu\nu}\hat{p}^{\nu},\\
\hat{p'}_{\mu} = \hat{p}_{\mu}+\frac{1}{2}\beta
\epsilon_{\mu\nu}\hat{x}^{\nu}.
\end{eqnarray}
Hamiltonian (\ref{50}) can now be written as
\begin{equation}\label{52}
\widehat{{\cal
H}}_{nc}=\frac{1}{4}\left(1-\omega^2\theta^2\right)\eta^{\mu\nu}\hat{p}_\mu
\hat{p}_\nu+\left(\omega^2-\frac{\beta^{2}}{16}\right)\eta^{\mu\nu}\hat{x}_\mu
\hat{x}_\nu-\left(\frac{\beta}{4}+\theta\omega^2\right)\epsilon^{\mu\nu}\hat{x}_\mu
\hat{p}_\nu,
\end{equation}
where the commutation relations (\ref{21}) are satisfied. The
annihilation operator is defined as
\begin{equation}\label{53}
a_\mu=\sqrt{\Xi}\hat{x}_\mu+\frac{i}{2\sqrt{\Xi}}\hat{p}_\mu,
\end{equation}
with
\begin{equation}\label{153}
\Xi=\left(\frac{\omega^2-\beta^2/16}{1-\theta^2\omega^2}\right)^{\frac{1}{2}},
\end{equation}
such that relations (\ref{23}) are satisfied. The Hamiltonian
(\ref{52}) can now be written in terms of $a$ and $a^\dag$
\begin{equation}\label{54}
\begin{array}{lll}
{\widehat{\cal H}}_{nc}
=\frac{\gamma}{2}\eta^{\mu\nu}\left(a_\mu^\dag a_\nu+a_\mu
a_\nu^\dag\right)-\frac{i\alpha}{2}\epsilon^{\mu\nu}\left(a_\mu
a_\nu^\dag-a_\nu a_\nu^\dag\right),\\
\end{array}
\end{equation}
where $\alpha$ and $\gamma$ are defined as
\begin{equation}\label{47}
\alpha=\left(\theta\omega^2+\frac{\beta}{4}\right)\hspace{3mm}\mbox{and}\hspace{3mm}
\gamma=\left[\left(1-\omega^2\theta^2\right)\left(\omega^2-\frac{\beta^2}{16}\right)\right]^{1/2}.
\end{equation}
Note that this Hamiltonian is Hermitian. Setting $\theta=0$ and
$\beta=0$, one recovers equation (\ref{25}).

As may be readily seen, the commutative solutions are not
solutions of the present case. To solve the WD equation,
${\widehat{\cal H}}_{nc} |\Psi \rangle = 0$, it seems natural to
follow the procedure used previously, \textit{i.e.} defining a
ground state and finding the rest of the solutions, if they exist,
by the action of an appropriate operator. To solve the WD
equation, the overall constant, $\gamma$, taken out from equation
(\ref{54}), does not have any effect and we therefore drop it
here. Now, Hamiltonian (\ref{54}) can be written as
\begin{equation}\label{542}
{\widehat{\cal H}}_{nc}=\left(a_1^\dag a_1-a_2^\dag a_2\right)+i
\vartheta \left(a_1 a_2^\dag-a_2 a_1^\dag\right),
\end{equation}
where $\vartheta=-{\alpha}/{\gamma}$. If one wants to build
solutions using the above procedure, one must define annihilation
and creation-like operators such that
\begin{equation}\label{543}
\left[{\widehat{\cal H}}_{nc},{\cal A}\right]=\left[{\widehat{\cal
H}}_{nc},{\cal A}^\dag\right]=0.
\end{equation}
Obviously if $|\Psi\rangle$ satisfies the WD equation then so do
${\cal A}|\Psi\rangle$ and ${\cal A}^\dag |\Psi\rangle$. To find
the above operators, we introduce the following annihilation-like
and creation-like operators (see Appendix)
\begin{eqnarray}\label{5411}
\verb"A"&=&\frac{1}{\sqrt{1+\vartheta^2}}\left[a_1 a_2-\frac{i \vartheta}{2}\left(a_1^2+a_2^2\right)\right],\\
\verb"A"^\dag&=&\frac{1}{\sqrt{1+\vartheta^2}}\left[a_1^\dag
a_2^\dag+\frac{i
\vartheta}{2}\left({a_1^\dag}^2+{a_2^\dag}^2\right)\right].
\end{eqnarray}
With a little algebraic calculations it can be shown that the
above operators satisfy the commutation relations (\ref{543}) and
\begin{eqnarray}
\left[\verb"A",\verb"A"^\dag\right]=1+\hat{n}_1+\hat{n}_2,
\label{eq-com}
\end{eqnarray}
where $\hat{n}_1$ and $\hat{n}_2$ are the number operators
corresponding to operators $a_1$ and $a_2$. Therefore, since the
ground state $|\verb"0"\rangle=|0,0\rangle$ is a solution of the
WD equation, ${\verb"A"^\dag}^n|\verb"0"\rangle$ becomes a
solution too. For a general solution represented by
$|2\verb"N"\rangle$, we have the following properties resulting
from equation (\ref{eq-com})
\begin{eqnarray}\label{5412}
\verb"A"|\verb"2N"\rangle&=&\verb"N"|\verb"2(N-1)"\rangle,\\
\verb"A"^\dag|\verb"2N"\rangle&=&\verb"(N+1)"|\verb"2(N+1)"\rangle.
\end{eqnarray}
\begin{figure}
\centerline{\begin{tabular}{ccc}
\epsfig{figure=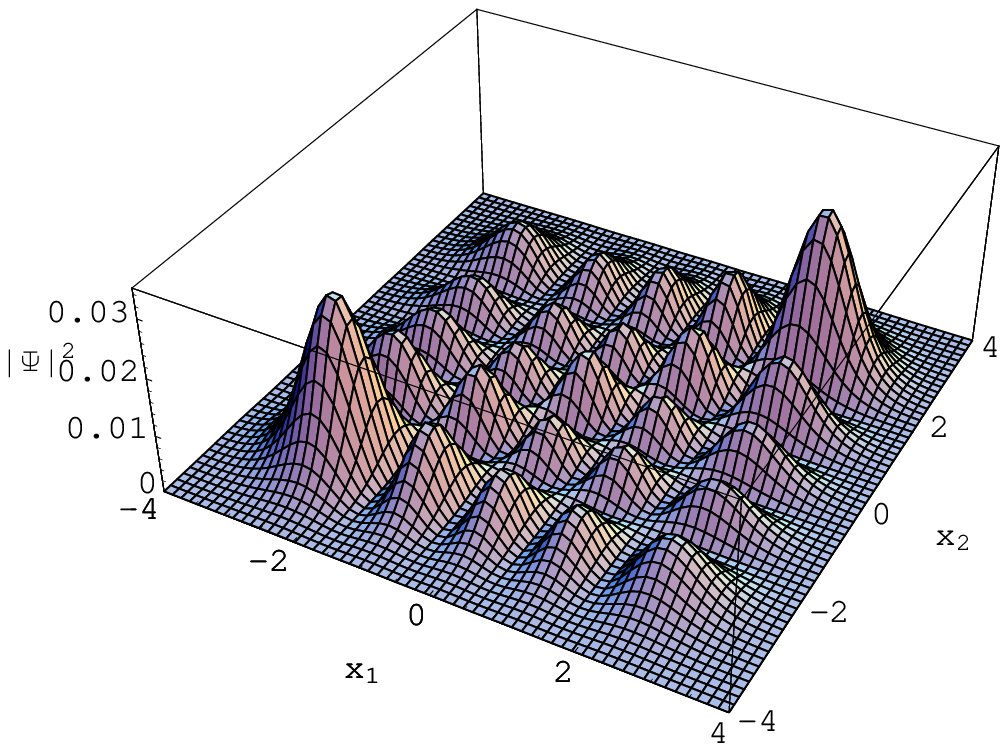,width=7cm}\hspace{1cm}
\epsfig{figure=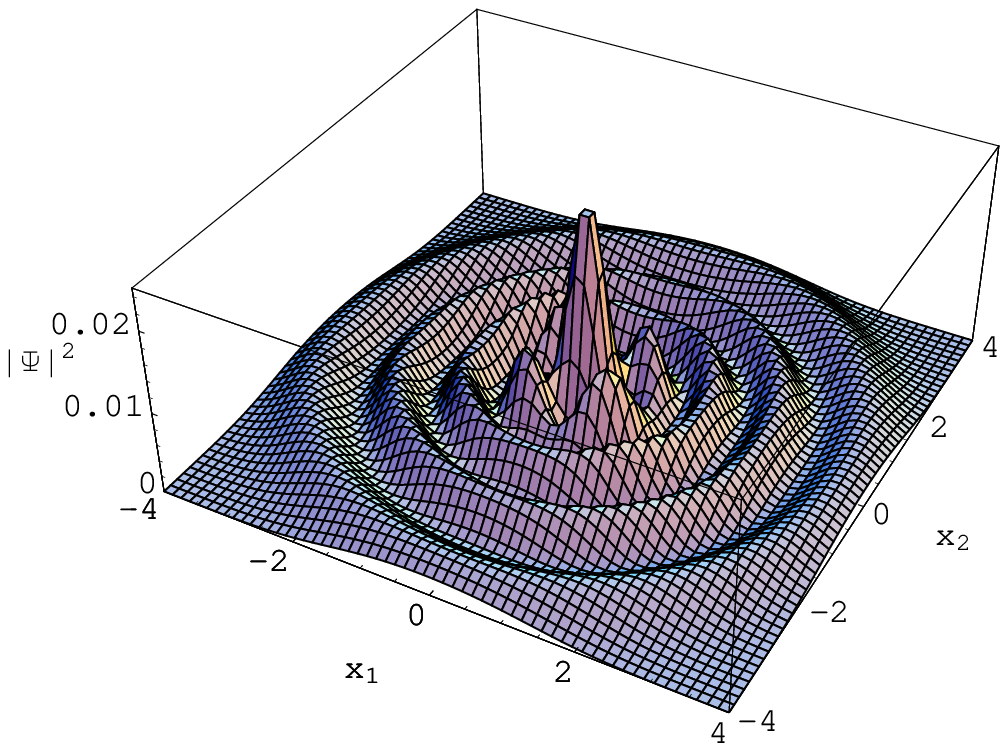,width=7cm}\\
\epsfig{figure=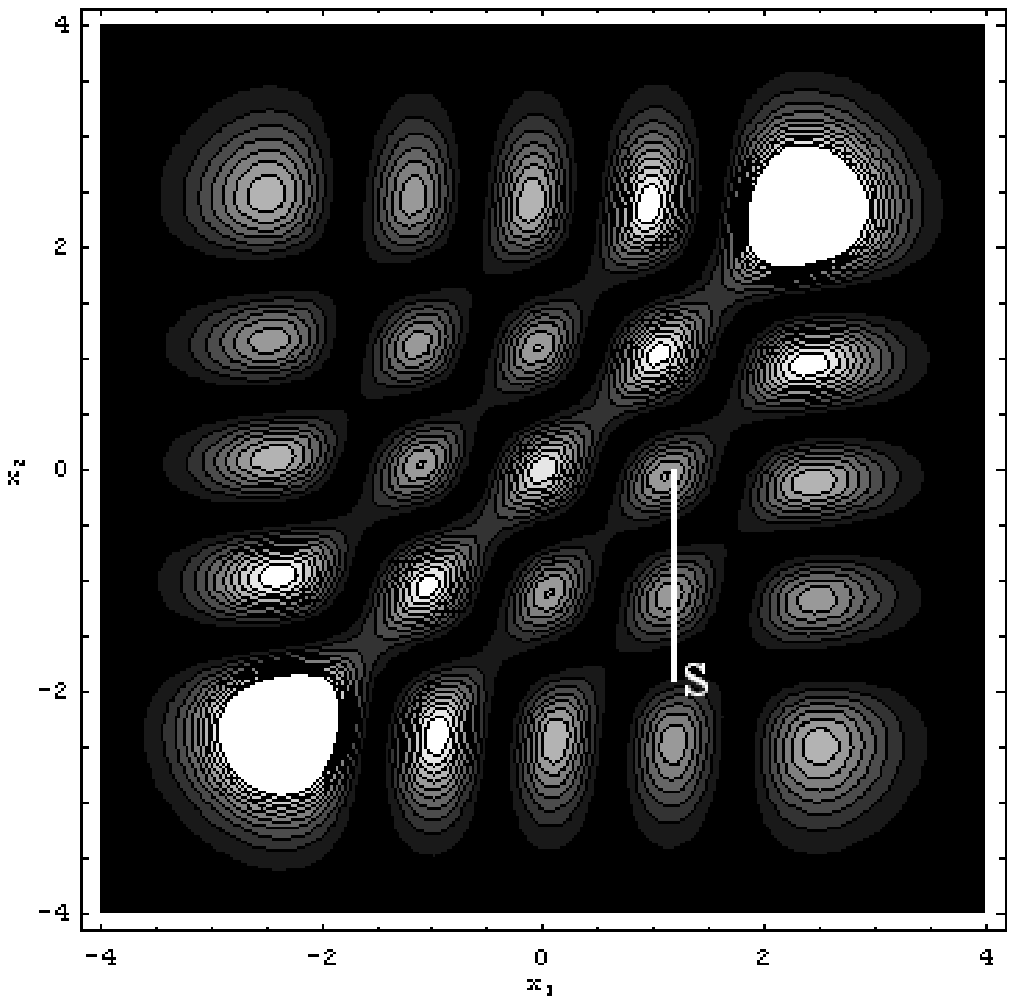,width=6cm}\hspace{2cm}
\epsfig{figure=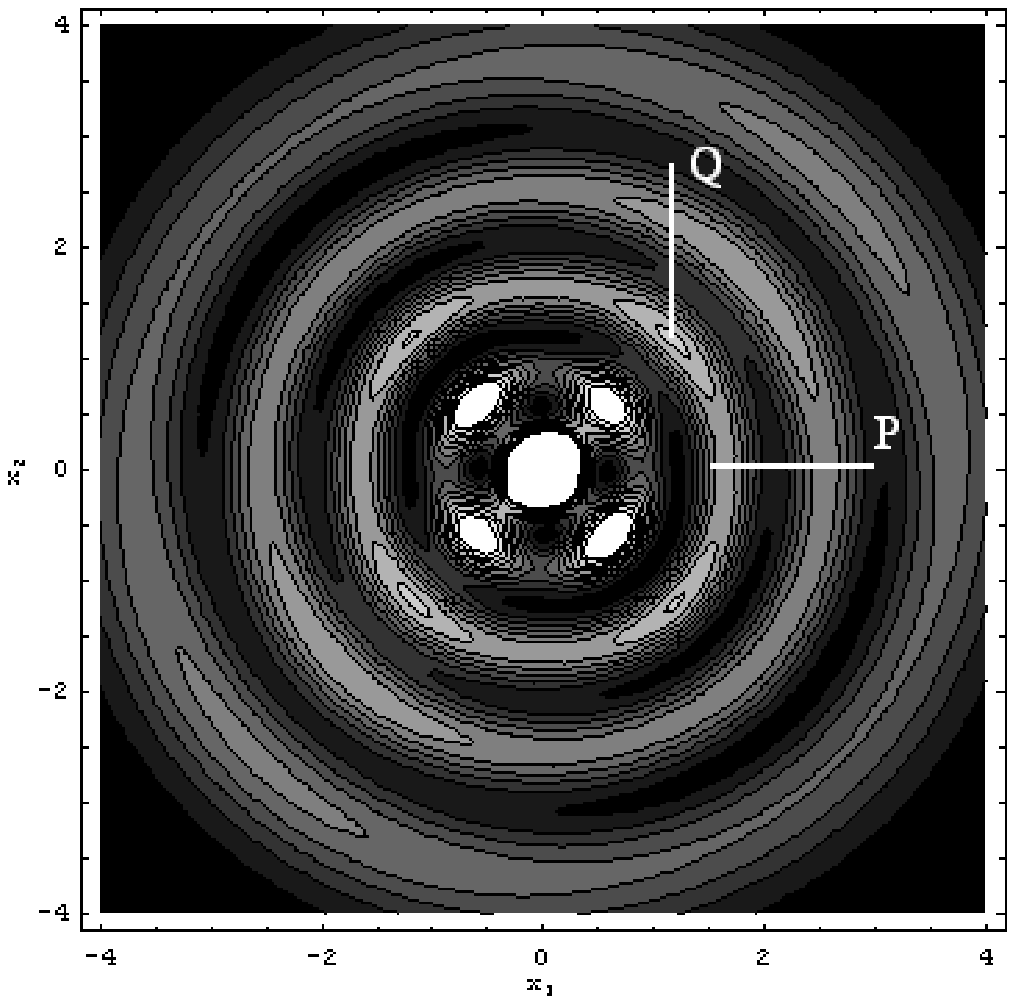,width=6cm}
\end{tabular}}
\caption{\footnotesize Left, the commutative and right, the
noncommutative square of the wavefunction. The corresponding
contour plots are drawn below each figure.} \label{fig2}
\end{figure}
These solutions have counterparts in the commutative case such
that if the parameter of noncommutativity, $\vartheta$, vanishes
$|\verb"2N"\rangle$ reduces to $|2n\rangle$ (for \verb"N"=n) which
is the solution for the commutative case represented by equation
(\ref{26}). We may then write
\begin{eqnarray}\label{5413}
|\verb"0"\rangle&=&|0,0\rangle,\\
|\verb"2"\rangle&=&\frac{1}{\sqrt{1+\vartheta^2}}
\left[|1,1\rangle+\frac{i\sqrt{2}\vartheta}{2}\left(|2,0\rangle+|0,2\rangle\right)\right],\\
|\verb"4"\rangle&=&\frac{1}{2(1+\vartheta^2)}\left[(2-\vartheta^2)|2,2\rangle+i
\vartheta
\sqrt{6}\left(|3,1\rangle+|1,3\rangle\right)-\frac{\vartheta^2
\sqrt{6}}{2}\left(|4,0\rangle+|0,4\rangle\right)\right],
\end{eqnarray}
and so on, where $|n,m\rangle$ is defined in (\ref{263}). We note
that setting $\vartheta=0$, the above solutions reduce to that of
the commutative ones. Also note that in this case
$|\verb"2N"\rangle$ is a mixture of the states represented by
$|n,m\rangle$ with $n+m=\verb"2N"$. In the $x$-representation the
states $|n,m\rangle$ should be replaced with their corresponding,
Hermite polynomials, that is
\begin{equation}\label{2624}
\Psi_{n,m}(x_1,x_2)={\cal M}
H_n(x_1)H_{m}(x_2)e^{-\omega(x_1^2+x_2^2)},
\end{equation}
where ${\cal M}$ is a normalization factor and the ket
$|\Psi\rangle$ can be written as
\begin{equation}\label{2625}
|\Psi\rangle=\sum_{\verb"N"}c_{\verb"N"}|\verb"2N"\rangle,
\end{equation}
where $c_{\verb"N"}$ is defined in (\ref{2641}). The corresponding
wave function and its contour plot is shown in figure 1.

A discussion on the quantum solutions obtained above would now be
in order. It is well known that the WD equation does not posses
time evolution, the so called problem of time in quantum
cosmology. The solutions presented above may offer an explanation
as to how to incorporate time evolution in these solutions. This
is particularly interesting in the present context since the
interpretation given below is compatible with the behavior of the
classical solutions presented in \cite{nima}.

Let us assume that an initial condition for the classical
solutions exist. One may then deduce the corresponding quantum
states. It would now be possible to provide an evolutionary
explanation of the wavefunctions by considerations of the
probability density defined in quantum mechanics. This can be
achieved by noting that a small perturbation from any given state
would, naturally, encourage this state to end up in another state
with the higher probability of existence. Such a transition in the
quantum state from low probability to a possible higher
probability may be viewed as an evolutionary process. Naturally,
the state corresponding to the local highest probability would
resist any perturbation it may experience in that it will not make
a transition to any other state\footnote{To be precise, it exists
around the peak.}. This can be seen from the plots in figure 1.
For suppose that the scale factors $a=R=1$. This is equivalent to
the choice $x_1=1$ and $x_2=0$, see equations (\ref{9}),
representing the points $S$ and $P$ in the contour plots. The
mechanism described above would prohibit the point $S$ to move
under any perturbation since it is already located at the point of
local highest probability. However, the point $P$ would move
towards the point $Q$ which has the highest probability amplitude
and provides an allowed transition from the point $P$. This point
corresponds to the state where the usual scale factor has grown
relative to the internal scale factor such that
$R\rightarrow\infty$. One may then conclude that noncommutativity
could provide for the dynamical compactification of the external
scale factor relative to the scale factor of the ordinary
universe. In addition, one may also conclude from the above
discussion that the scale factor representing the extra dimensions
could oscillate about a mean value corresponding to a point with
local highest probability (see the footnote 1).

An example of interest from the string theory point of view is the
case $d=6$ for which one can write
\begin{equation}\label{scalefactorssix}
\begin{array}{ll}
a=\left(x_1+x_2\right)^{-\frac{1}{18}}\left(x_1-x_2\right)^{\frac{1}{6}},\\
R=\left(x_1+x_2\right)\left(x_1-x_2\right)^{-\frac{5}{3}}.
\end{array}
\end{equation}
For the noncommutative case the transition from the point $P$ to
the point $Q$ is equivalent to the transition from $x_1=1$ and
$x_2=0$ to $x_1=x_2$. Therefore, during this transition process
the value of $\epsilon=x_1-x_2$ becomes progressively smaller,
making the second terms in the scale factors
(\ref{scalefactorssix}) the dominant player, since the scale
factor of extra dimensions $a$ goes to zero and $R$ goes to
infinity when $\epsilon\rightarrow0$. This means that the
compactification of extra dimensions relative to the scale factor
of the ordinary universe has been achieved during such
transition.\footnote{The above calculations and conclusions do not
depend to the choice $d=6$, except when $d<3$ which makes $R$
compactified relative to $a$.}

\section{Conclusions}
In this paper we have introduced the notion of noncommutativity
between the scale factors of the ordinary universe and extra
dimensions in a multidimensional cosmological model.

The quantum states of the universe were derived by solving the
corresponding WD equation. These solutions were generated by
defining certain creation and annihilation-like operators. To
discuss the physical properties of these quantum states and their
relation to the classical solutions, a mechanism was suggested for
the assignment of an evolutionary interpretation to these
solutions. This mechanism is based on the interpretation one may
infer from the behavior of the wavefunction when a small
external\footnote{Certainly, in quantum cosmology, the universe is
considered as one whole \cite{20} and the introduction of an
external force is irrelevant. But, because of the lack of a full
theory to describe the universe, these small external forces are
merely used to afford a better understanding of the discussions
presented here.} perturbation is imposed and the transition of any
given state to an allowed state with a higher probability is
considered. Note that the source of this external perturbation is
unknown in the absence of a theory describing the quantum state of
the universe where it can be treated as one whole. An evolutionary
process was hence emerged, upon whose  application to the
wavefunction, the same properties regarding stabilization and
compactification of the extra dimensions, as those of the
classical solutions, appeared \cite{nima}.

\appendix
\section{Appendix}
In this appendix we consider the construction of the annihilation
and creation-like operators discussed above and used to obtain the
solutions of the WD equation. Let the general form for these
operators take the form
\begin{equation}\label{26251}
{\cal{A}}_m=\sum^{m}_{i=0} h_i(\vartheta) a_1^i a_2^{m-i} ,
\end{equation}
where $n=1,2,3,...$  and $h_i(\vartheta)$ are some functions such
that relation (\ref{543}) is satisfied. One therefore starts to
build the solutions of the WD equation by operating $k$-times with
the creation-like operator ${\cal{A}}^{\dag}_m$ on the ground
state $|0,0\rangle$. For example, for $k=1$ we write
\begin{equation}\label{26259}
{{\cal{A}}^{\dag}_m} |0,0\rangle=\left[\sum^{m}_{i=0}
h^*_i(\vartheta) {a^{\dag}_1}^i
{a^{\dag}_2}^{m-i}\right]|0,0\rangle=\sum^{m}_{i=0}h^*_i(\vartheta)
\sqrt{i!(n-i)!} |i,m-i\rangle.
\end{equation}
We are interested in solutions that have commutative counterparts,
\textit{i.e.}  solution having the property that when
$\vartheta\rightarrow0$ they reduce to the commutative solutions.
Obviously, in the commutative case, solutions of the WD equation
(\ref{26}) have the form $|n,n\rangle$. Therefore, setting
$\vartheta=0$, solutions (\ref{26259}) and their commutative
counterparts impose
\begin{eqnarray}\label{2625111}
m&=&2n,\\
h^*_{i\neq n}(\vartheta\rightarrow0)&=&0,\\
h^*_n(\vartheta\rightarrow0)&=&C,
\end{eqnarray}
where $C$ is a constant. Now let $m=2n=4$. In this case the
corresponding commutative solutions are
\begin{equation}\label{1111111111}
 |0,0\rangle ,\hspace{.5cm} |2,2\rangle , \hspace{.5cm}|4,4\rangle, \hspace{.5cm}
 \cdots
 \end{equation}
Obviously, these solutions do not represent the complete spectrum
of the commutative solutions; $|1,1\rangle$ is not present for
example. Note that this is true for all $m=2n>2$. These solutions
do not seem to be interesting since the corresponding commutative
solutions are not complete, hence we have ignored them. The only
choice remaining is therefore $m=2n=2$, discussed above. It is
worth noting that the coefficients $h_i(\vartheta)$ have already
been calculated and may be read off from equation (\ref{543}).

\end{document}